\documentstyle[12pt,fleqn]{article}

\font\twlgot =eufm10 scaled \magstep1
\font\egtgot =eufm8
\font\sevgot =eufm7

\font\twlmsb =msbm10 scaled \magstep1
\font\egtmsb =msbm8
\font\sevmsb =msbm7

\newfam\gotfam
\def\pgot{\fam\gotfam\twlgot}
\textfont\gotfam\twlgot
\scriptfont\gotfam\egtgot
\scriptscriptfont\gotfam\sevgot
\def\got{\protect\pgot}

\newfam\msbfam
\textfont\msbfam\twlmsb
\scriptfont\msbfam\egtmsb
\scriptscriptfont\msbfam\sevmsb
\def\Bbb{\protect\pBbb}
\def\pBbb{\relax\ifmmode\expandafter\Bb\else\typeout{You cann't use
Bbb in text mode}\fi}
\def\Bb #1{{\fam\msbfam\relax#1}}

\newcommand{\gF}{{\got F}}

\def\thebibliography#1{\section*{References}\list
  {[\arabic{enumi}]}{\settowidth\labelwidth{#1}\leftmargin\labelwidth
    \advance\leftmargin\labelsep
    \usecounter{enumi}}
    \def\newblock{\hskip .11em plus .33em minus .07em}
    \sloppy\clubpenalty4000\widowpenalty4000
    \sfcode`\.=1000\relax}

\let\Large=\large

\def\op#1{\mathop{\fam0 #1}\limits}

\newcommand{\id}{{\rm Id\,}}

\newcommand{\di}{{\rm dim\,}}

\newcommand{\beq}{\begin{equation}}
\newcommand{\eeq}{\end{equation}}
\newcommand{\ben}{\begin{eqnarray}}
\newcommand{\een}{\end{eqnarray}}
\newcommand{\be}{\begin{eqnarray*}}
\newcommand{\ee}{\end{eqnarray*}}
\newcommand{\bea}{\begin{eqalph}}
\newcommand{\eea}{\end{eqalph}}

\newcommand{\cO}{{\cal O}}
\newcommand{\cF}{{\cal F}}

\newcommand{\al}{\alpha}

\newcommand{\f}{\phi}

\newcommand{\m}{\mu}

\newcommand{\G}{\Gamma}
\newcommand{\th}{\theta}

\newcommand{\w}{\wedge}
\newcommand{\wt}{\widetilde}

\newcommand{\dr}{\partial}
\newcommand{\ar}{\op\longrightarrow}
\newcommand{\ot}{\otimes}

\newcounter{eqalph}
\newcounter{equationa}
\newcounter{theorem}
\newcounter{proposition}
\newcounter{lemma}
\newcounter{corollary}
\newcounter{definition}

\def\thedefinition{\arabic{definition}}

\newenvironment{eqalph}{\stepcounter{equation}
\setcounter{equationa}{\value{equation}}
\setcounter{equation}{0}

\begin{eqnarray}}{\end{eqnarray}\setcounter{equation}{\value{equationa}}}

\newcommand{\mar}[1]{}

\begin{document}
\hbox{}

{\parindent=0pt 

{ \Large \bf Differential geometry of fibre bundles over
foliated manifolds}
\bigskip

{\sc G.SARDANASHVILY}

{ \small

{\it Department of Theoretical Physics,
Physics Faculty, Moscow State University, 117234 Moscow, Russia

E-mail: sard@grav.phys.msu.su}
\bigskip

{\bf Abstract.} 
Any leafwise connection on a fibre bundle over a foliated manifold
is proved to come from a connection on this fibre bundle.
\bigskip

{\bf Mathematics Subject Classification (2000):} 53C05, 53C12

} }

\bigskip\bigskip

Let $(Z,\cF)$ be a foliated manifold, where $\cF$ denotes a regular
foliation of $Z$. Let $Y\to Z$ be a fibre bundle over $Z$. 
A leafwise connection on $Y\to Z$ is defined as a partial connection
along the leaves of a folition $\cF$. Obviously, each connection on
$Y\to Z$ yields a leafwise connection. We show that any leafwise connection
on $Y\to Z$ is of this type. This fact is important, e.g., for geometric
quantization of symplectic foliations.

We start from the notion of a leafwise differential calculus on a
foliated manifold. 
All manifolds throughout are smooth,
real, finite-dimensional, Hausdorff,
paracompact, and connected.

If $(Z,\cF)$ is a (regular) foliated manifold, 
there is an adapted
coordinate atlas
\mar{spr850}\beq
\Psi_\cF=\{(U;z^\al; z^A)\},\quad \al=1,\ldots,\di\cF, 
\quad A=\di\cF+1,\ldots,\di Z,
\label{spr850}
\eeq
of $Z$ such that,
for every leaf $F$ of a foliation $\cF$, the connected components
of $F\cap U$ are described by the equations
$z^A=$const.,
and transition functions of coordinates $z^A$ are independent of the 
remaining ones \cite{rei}. 

The
tangent spaces to leaves of a foliation $\cF$ assemble into 
an involutive distribution $T\cF\to Z$ on $Z$ called the tangent
bundle to $\cF$. Its global sections
constitute the real Lie algebra $T\cF(Z)$
of vector fields on $Z$ subordinate to $T\cF$.
This Lie algebra is a left $C^\infty(Z)$-submodule of the
derivation module of the $\Bbb R$-ring
$C^\infty(Z)$ of smooth real functions on $Z$. Let $S_\cF(Z)$ be the kernel
of $T\cF(Z)$, acting on $C^\infty(Z)$. It consists of functions, constant
on leaves of $\cF$ (i.e., foliated functions in the terminology of
\cite{vais73}). Then 
$T\cF(Z)$ is the Lie $S_\cF(Z)$-algebra of derivations of $C^\infty(Z)$,
seen as a $S_\cF(Z)$-ring.
Let us consider the 
Chevalley--Eilenberg differential calculus
over the $S_\cF(Z)$-ring $C^\infty(Z)$. It is defined as a 
cochain subcomplex
\mar{spr892}\beq
0\to S_\cF(Z)\ar C^\infty(Z)\ar^{\wt d} \gF^1(Z) \ar^{\wt d}\cdots
\ar^{\wt d} \gF^{\di\cF}(Z) \to 0 \label{spr892}
\eeq
of the Chevalleqy--Eilenberg complex of
the Lie $S_\cF(Z)$-algebra  
with coefficients in the ring $C^\infty(Z)$, where $\gF^r(Z)$, $r>0$,
is a $C^\infty(Z)$-module of
$C^\infty(Z)$-multilinear skew-symmetric maps
\be
\f:\op\times^r T\cF(Z) \to C^\infty(Z)
\ee
\cite{book00}.
These maps are precisely global sections of the exterior product
$\op\w^r T\cF^*$ of the dual $T\cF^*\to Z$ of $T\cF\to Z$. 
They are called the leafwise forms
on a foliated manifold $(Z,\cF)$ (one should distinguish them from
foliated forms in \cite{vais73}
which are exterior forms on $Z$ constant on leaves of a foliation).
With respect to the adapted coordinates $(z^\al;z^A)$
(\ref{spr850}), the leafwise forms read
\mar{spr890}\beq
\f=\frac1{r!}\f_{\al_1\ldots \al_r}\wt dz^{\al_1}\w\cdots\w \wt dz^{\al_r},
\label{spr890}
\eeq
where $\{\wt dz^\al\}$ are the duals of the holonomic fibre
bases $\{\dr_\al\}$ for $T\cF$. 
The Chevalley--Eilenberg coboundary operator on leafwise forms
is given by the expression
\mar{spr891}\beq
\wt d\f= \wt dz^\m\w \dr_\m\f=\frac{1}{r!}
\dr_\m\f_{\al_1\ldots \al_r}\wt dz^\m\w\wt dz^{\al_1}\w\cdots\wt dz^{\al_r},
\qquad \m,\al\leq \di\cF, \label{spr891}
\eeq
(it is precisely the operator $d_f$ in \cite{vais73}).
It is called  the leafwise differential,
and $(\gF^*(Z),\wt d)$ is the 
leafwise differential
calculus \cite{hect}.

Let $V\cF=TZ/T\cF\to Z$ be the normal bundle to a foliation $\cF$ and $V\cF^*$
the dual of $V\cF$. There are the exact sequences of vector bundles
\mar{spr916,7}\ben
&& 0\to T\cF\ar^i_Z TZ\ar^j_Z V\cF \to 0, \label{spr916}\\
&& 0\to V\cF^*\ar^{j^*}_Z T^*Z\ar^{i^*}_Z T\cF^* \to 0. \label{spr917}
\een
The epimorphism $i^*$ in (\ref{spr917}) provides an order-preserving 
epimorphism of the
graded differential algebra $(\cO^*(Z),d)$ of exterior forms on $Z$ to
the algebra $(\cF^*(Z),\wt d)$ of leafwise forms such that $i^*\circ d=\wt
d\circ i^*$. With respect to the adapted coordinates
(\ref{spr850}), this epimorphism reads
\be
i^*:dz^\al \mapsto \wt dz^\al, \qquad i^*: dz^A\to 0.
\ee

Turn now to the notion of a leafwise connection on a fibre bundle
\mar{lmp17}\beq
\pi:Y\to Z \label{lmp17}
\eeq
over a foliated manifold
$(Z,\cF)$. The inverse
images $\pi^{-1}(F)$ of leaves $F$ of the foliation $\cF$ of $Z$ provides
a (regular) foliation $Y_\cF$ of $Y$.  
One can always choose an adapted coordinate atlas $\{(U;z^\al; z^A)\}$
(\ref{spr850}) of a foliated manifold $(Z,\cF)$ such that $U$ are
trivialization domains of a fibre bundle $Y\to Z$. Let $(z^\al;
z^A;y^i)$ be the 
corresponding bundle coordinates on $Y\to Z$. They are also adapted
coordinates on the foliated manifold $(Y,Y_\cF)$. 

Let us note that, in the terminology of \cite{vais73}, a fibre bundle $Y\to Z$
(\ref{lmp17}) is called foliated if its transition functions are
constant along leaves of $\cF$.

Given the
tangent bundle $TY_\cF\to Y$ to the foliation $Y_\cF$ of $Y$, we have the
exact sequence of vector bundles
\mar{lmp18}\beq
0\to VY\ar_Y TY_\cF\ar_Y Y\op\times_Z T\cF\to 0, \label{lmp18}
\eeq
where $VY$ is the vertical tangent bundle of $Y\to Z$. Its splitting exists,
and is called a leafwise connection on a fibre bundle $Y\to Z$. It is
represented by a $TY_\cF$-valued leafwise
one-form
\mar{lmp21}\beq
A_\cF=\wt dz^\al\ot(\dr_\al +A^i_\al\dr_i). \label{lmp21}
\eeq

Let $J^1_\cF Y$ be the subbundle 
of the 
tensor product $T\cF^*\op\ot_YT Y_\cF\to Y$
whose image under the surjection
\be
T\cF^*\op\ot_YT Y_\cF\ar_Y Y\op\times_Z (T\cF^*\op\ot_Z T\cF) 
\ee
coincides with the canonical section $\th_\cF(Y)=\wt d z^\al\ot \dr_\al$
of the pull-back 
\be
Y\op\times_Z (T\cF^*\ot T\cF)\to Y.
\ee
One can think of $J^1_\cF Y$ as being a leafwise jet manifold of a fibre
bundle $Y\to Z$. With respect to the adapted bundle coordinates
$(z^\al;z^A;y^i)$, its elements read
\be
\wt dz^\al\ot(\dr_\al +y^i_\al\dr_i).
\ee
In particular, every section $s$ of a fibre bundle $Y\to Z$
gives rise to the section  
\be
J^1_\cF s=\wt dz^\al\ot(\dr_\al +\dr_\al s^i\dr_i)
\ee
of the leafwise jet bundle $J^1_\cF Y\to Z$.

Similalry to the case of a connection, one can show that
every leafwise connection on a fibre bundle $Y\to Z$ is a global section
of the affine leafwise jet bundle 
\be
\pi^1_0:J^1_\cF Y\to Y,
\ee
 and {\it vice versa}.
It follows that leafwise connections constitute an affine space
modelled on the vector space of sections of the vector bundle
\mar{lmp22}\beq
T\cF^*\op\ot_YVY\to Y. \label{lmp22}
\eeq
In particular, we have the 
leafwise covariant differential 
\be
&& \nabla_\cF= \id -A_\cF\circ\pi^1_0 : J^1_\cF Y\to T\cF^*\op\ot_YVY, \\
&& \nabla_\cF s=(\wt ds^i - (A_\al^i\circ s) \wt dz^\al)\ot\dr_i.
\ee

Since the exact sequence (\ref{lmp18}) is a subsequence of the exact sequence
\be
0\to VY\ar_Y TY\ar_Y Y\op\times_Z TZ\to 0,
\ee
every connection 
\be
\G=dz^A\ot(\dr_A + \G^i_A\dr_i) + 
dz^\al\ot(\dr_\al +\G^i_\al\dr_i)
\ee
on a fibre bundle $Y\to Z$ yields a leafwise connection
\mar{lmp23}\beq
\G_\cF=\wt dz^\al\ot(\dr_\al +\G^i_\al\dr_i). \label{lmp23}
\eeq

Conversely, let $A_\cF$ (\ref{lmp18}) be a leafwise connection on a
fibre bundle $Y\to Z$ and $\G_\cF$
(\ref{lmp23}) a leafwise connection which comes from a connection on
$Y\to Z$. Their affine difference over $Y$ is a section
\be
Q=A_\cF-\G_\cF=\wt dz^\al\ot(A^i_\al -\G^i_\al)\dr_i
\ee
of the vector bundle (\ref{lmp22}). Given some splitting 
\mar{lmp25}\beq
B: \wt dz^\al \mapsto dz^\al- B^\al_A dz^A \label{lmp25}
\eeq
of the exact sequence (\ref{spr917}), the composition
\be
&& B\circ Q: Y\to T^*Z\op\ot_Y VY,\\
&& B\circ Q=(dz^\al- B^\al_A dz^A)\ot(A^i_\al -\G^i_\al)\dr_i,
\ee
is a soldering form on a fibre bundle $Y$. Then 
\be
\G+B\circ Q= 
dz^A\ot(\dr_A + \G^i_A\dr_i -B^\al_A (A^i_\al -\G^i_\al)\dr_i) + 
dz^\al\ot(\dr_\al +A^i_\al\dr_i)
\ee
is a desired connection on a fibre bundle $Y\to Z$ which yields the leafwise 
connection $A_\cF$ (\ref{lmp18}).

Of course, such a connection is not unique, but depends on the choice
of a trivialization $B$ (\ref{lmp25}) of the exact sequence (\ref{spr917}).

\end{document}